\documentclass[12pt]{article}
\usepackage{euscript,amssymb}

\title{Restrictions on dilatonic brane-world models}
\author{ A. Zhuk\thanks{e-mail: zhuk@paco.net}\\ Department of
Physics, University of Odessa,\\ 2 Dvoryanskaya St., Odessa 65100,
Ukraine }

\date{}
\begin{document}
\maketitle
\abstract{ We consider dilatonic brane world models with a
non-minimal coupling between a dilaton and usual matter on a
brane. We demonstrate that variation of the fundamental constants
on the brane due to such interaction leads to strong restrictions
on parameters of models. In particular, the experimental bounds on
variation of the fine structure constant rule out non-minimal
dilatonic models with a Liouville-type coupling potential
$f(\varphi ) = \exp (b \varphi )$ where $b \sim \mathcal{O}(1)$.


\vspace{2cm}


%
%
%
Brane-world models have been the subject of intensive
investigation for the last few years. They offer an interesting
alternative (with respect to the Kaluza-Klein model) to the
standard multidimensional gravity and cosmology. The main feature
of this approach consists in a proposal where the standard matter
(SM) fields are localized on the brane (4-dimensional hypersurface
which corresponds to our Universe) whereas the gravitational field
can propagate in the full multidimensional space-time. It sheds a
new light on the problem of the large hierarchy and leads to new
designing properties and phenomena for multidimensional models.
Thus, it is important to predict observable effects which can
confirm
such brane-world approach.

Obviously,
SM particles may escape from the brane
into a bulk resulting in the violation of the energy-momentum and
charge conservation laws in the brane \cite{DRT1}. Such effect can
take place, for example, if SM particles interact with bulk
fields. A lot of papers were devoted to the problem of the
interaction between radions and SM fields (see \cite{CGRT,GRW} and
references therein). Radions usually describe relative motion of
the branes. For realistic models, it is usually supposed that
there is a mechanism for the brane stabilization with respect to
each other. Let $b_0$ be the scale of stabilization of the
inter-brane distance and $\psi (x)$ the small fluctuations
(radions) around it. Then, an induced 4-D metric on the brane
located in an additional dimension at $y= y_0$ reads:
$h_{\mu\nu}(x,y_0) = A_0 \exp (c_0\psi (x))\tilde h_{\mu\nu}(x)$,
where $A_0$ is a dimensionless warp factor corresponding to the
scale of stabilization $b_0$ and $c_0 \sim 1/M_{\mathrm{EW}}$ (in
the ADD (Arkani-Hamed, Dimopoulos, Dvali) brane approach $c_0 \sim
1/M_{\mathrm{Pl}}$ \cite{GRW,ADD_astr}). Let $\Phi (x)$ represents
a matter field (SM) on the $D_0$-dimensional brane with a
Lagrangian $L = L (\Phi (x), h (x,y_0))$ and following action
\begin{equation}\label{1} S = \int d^{D_0}x \sqrt{|h|} L (\Phi (x),
h (x,y_0))\, .\end{equation}
The corresponding Lagrangian density of the interaction between
radions and field $\Phi$ is
\begin{equation}\label{2} \mathcal{L}_{int} = \left.  \psi \frac{\delta
\mathcal{L}}{\delta \psi}\right|_{\psi = 0} =   \psi \left(
\frac{\delta \mathcal{L}}{\delta h^{\mu\nu}}\frac{\delta
h^{\mu\nu}}{\delta \psi}\right)_{\psi = 0} = - (c_0/2)\psi \left(
\sqrt{|h|}\, T^{\mu}_{\mu}\right)_{\psi=0}\, , \end{equation}
where $T^{\mu}_{\mu}$ is a trace of the energy-momentum tensor for
the Lagrangian $L$ with respect to the metric $h_{\mu\nu}$:
\begin{equation}\label{2a}
T_{\mu\nu} = -2 \frac{\delta L }{ \delta h^{\mu\nu}} + h_{\mu\nu}
L \Leftrightarrow \sqrt{|h|}T_{\mu\nu} = 2 \frac{\delta
\mathcal{L}}{ \delta h^{\mu\nu}}\, ,\;  \mathcal{L} = \sqrt{|h|}
L\, .
\end{equation}
Thus, the interaction between radions and SM field is absent for
fields with vanishing $T^{\mu}_{\mu}$, e.g. for massless fermions
and massless gauge bosons which are the quanta present in
high-energy experiments. By this reason graviscalars were
neglected in colliding experiments for studding of the brane-world
physics\footnote{We should emphasize that in the standard
Kaluza-Klein model the interaction (with $c_0 \sim
1/M_{\mathrm{Pl}}$) between graviscalars (gravexcitons \cite{GZ})
and massless particles is possible at tree level \cite{GZ2}.}.

Nevertheless, massless SM particles on the brane can interact at
tree level with other bulk fields, e.g. with a non-minimal dilaton
field. Moreover, as this scalar field lives in full 5-D space-time
(in the bulk), the coupling constant is $c_0 \sim 1/M$ (if 5-D
gravitational constant $\kappa^2_5 = M^{-3}$). Such interaction
may play an important role in the brane-world physics. Thus, it is
of interest to predict observable effects following from this type
of interaction and to obtain experimental restrictions on
parameters of the models.
There is an extensive list of papers devoted to the investigation
of the dilatonic brane-world models with a slightly different form
of the action (e.g.
\cite{CR,Lidsey,OS,ADKS,KSS,Youm,HLZ,GNS,BCG,MW,MB,NOOT,BDB,LM,Davis}).
They naturally follows from a low-energy limit of string theories
and have a dilatonic bulk potential and a dilatonic coupling
potential of the form of the Liouville potential
\cite{CR,Lidsey,OS,GNS,BDB}. For example, the action describing
non-minimal coupling between dilaton and a brane matter (in the
model with one brane) can be written in the form
\begin{eqnarray}\label{2b} S &=& \frac{1}{2\kappa^2_5} \int_{M_5} d^5 x
\sqrt{|\bar g|} e^{-2\varphi} \{ R[\bar g] + 4\bar
g^{ab}\partial_a \varphi
\partial_b \varphi - 2\kappa^2_5 \; V \} + \\ &+& \int_{M_4}
d^4 x \sqrt{|\bar h|} e^{-2\varphi} \{ -T + L_m [\bar h] \} \, ,
\nonumber
\end{eqnarray}
where $M_5$ is the 5-D manifold with metric $\bar g_{ab}$ ($a,b =
0,1,2,3,5$) and the 4-D hypersurface $M_4$ is the brane with
induced metric\footnote{For simplicity, we consider the case of
one brane located at the additional coordinate $y = y_0$. Let
$n_a$ be a unite space-like vector normal to the brane. Then, the
induced metric on the brane is $\bar h_{ab} = \bar g_{ab} - n_a
n_b$. We also suppose that all space-time can be covered by the
normal Gauss coordinates where $n_a = n^a = (0,0,0,0,1)$. In this
case $\bar h_{a5}=\bar h^a_5=\bar h^5_a=0$ and $\bar h^{\mu}_{\nu}
= \delta^{\mu}_{\nu}$. These simplifications do not affect the
results of our paper.} $\bar h_{\mu \nu}$ ($\mu , \nu = 0,1,2,3$).
$T$ is a tension of the brane.
The Lagrangian $L_m[\bar h]$, constructed with the help of the
metric $\bar h$, corresponds to SM fields on the brane. 5-D
gravitational constant $\kappa^2_5$ is connected with 5-D
fundamental mass as follows: $\kappa^2_5 = M^{-3}$ and we usually
suppose\footnote{This assumption is valid for 5--D brane-world
models of the Randall--Sundrum--type \cite{RS}, where gravity is
effectively 4--dimensional below a length scale of the order
$M^2_{\mathrm{Pl}}/M^3_{\mathrm{EW}}$.
In the case of 5--D ADD approach, gravity becomes 5--dimensional
below $M^2_{\mathrm{Pl}}/M^3$. For such ADD models, to be in
accordance with gravitational experiments (see e.g.
\cite{experiment1}), mass scale $M$ should satisfy the following
restriction: $M \gtrsim 10^8$ GeV.} $M = M_{\mathrm{EW}} \sim
1TeV$. The dilatonic field $\varphi$ is dimensionless. Its
dimensions are restored with the help of the 5-D fundamental mass
$M$: $\varphi = M^{-3/2}\bar \phi = M^{-1}\phi$ where $\bar \phi$
and $\phi$ have dimensions of $\mathcal{O}(m^{3/2})$ and
$\mathcal{O}(m)$ (m is a unit of mass), respectively. A scalar
field $\phi$ has usual dimensions for scalar fields in 4-D
space-time (cf. \cite{LM}). The bulk potential $V$ (which we
consider as the bulk cosmological constant) has dimensions
$\mathcal{O}(m^5)$.
The tension $T$ and the matter Lagrangian $L_m$ have dimensions
$\mathcal{O}(m^4)$.

Action (\ref{2b}) is written in the string frame, where we suppose
that 5--D original metric $\bar g_{ab}$ and its induced metric
$\bar h_{\mu \nu}$ do not depend on the dilaton field $\varphi$.
Conformal transformation to the Einstein frame
\begin{equation}\label{2c} g_{ab} = \Omega^2_1 (\varphi ) \bar g_{ab} \equiv
e^{-(4/3) \varphi} \bar g_{ab} \end{equation}
yields
\begin{eqnarray}\label{3} S &=& \frac{1}{2\kappa^2_5} \int_{M_5} d^5 x
\sqrt{|g|} \{ R[g] - \frac43 g^{ab}\partial_a \varphi \partial_b
\varphi - 2\kappa^2_5\; e^{(4/3) \varphi}V \} + \\ &+& \int_{M_4}
d^4 x \sqrt{|h|} \{ -e^{(2/3) \varphi} T + e^{(2/3) \varphi}L_m
[\varphi , h] \} \, , \nonumber
\end{eqnarray}
where  $L_m [\bar h] = L_m [\Omega^{-2}h] \equiv L_m [\varphi ,
h]$.

It is clear that a non-minimal interaction of the dilatonic field
with SM fields can result in violation of the matter conservation
on the brane (see footnote 5 below). To be in accordance with
observations, $\varphi$ should be stabilized on the brane near
some value $\varphi_0$ or slightly vary during the Universe
evolution (at least from the time of nucleosynthesis). To estimate
a possible restrictions on the rate of such variations, it is
necessary to investigate a Lagrangian of interaction between
dilaton and SM fields.   Let $\varphi_0$ is the present value of
$\varphi$ and $\eta = M^{-1}\psi$ are small fluctuations around
it. Additionally, we slightly generalize our model to an arbitrary
dilaton coupling potential: $e^{-2 \varphi} \to f(\varphi)$
supposing only that function $f (\varphi)$ (as well as $\Omega_1
(\varphi )$) is the Liouville-type potential. It is of interest to
compare the Lagrangians of interaction in different frames.

\vspace{0.5cm}

1. String frame.

\vspace{0.5cm}

Here, the action of the SM fields on the brane and the Lagrangian
density of interaction read, correspondingly:
\begin{equation}\label{3a}
S_m = \int_{M_4} d^4 x \sqrt{|\bar h|} f(\varphi ) L_m [\bar h] \}
\, ,
\end{equation}
\begin{equation}\label{3b}
\mathcal{L}_{int} = \eta \left. \frac{\delta\left(\sqrt{\bar h}
f(\varphi ) L_m [\bar h] \right)}{\delta \varphi}\right|_{\varphi
= \varphi_0} = \eta \left. \frac{d f}{d \varphi} \sqrt{\bar h} L_m
[\bar h] \right|_{\varphi = \varphi_0}\, .
\end{equation}

\vspace{0.5cm}

2. Einstein frame.

\vspace{0.5cm}

In this frame, the SM action can be written as follows:
\begin{equation}\label{3c} S_m = \int_{M_4} d^4 x \sqrt{| h|}
\Omega^{-4}_1 (\varphi ) f(\varphi ) L_m [\varphi , h]  \equiv
\int_{M_4} d^4 x \sqrt{| h|} F(\varphi ) L_m [h] \} \, ,
\end{equation}
where $F(\varphi ) \equiv \Omega^{-4}_1 (\varphi ) f(\varphi ) f_1
(\varphi )$ and we imply $ L_m [\varphi , h] \equiv L_m
[\Omega^{-2}_1h] = f_1(\varphi ) L_m [h]$. The latter equality
(resulting in factorization) usually takes place for massless
particles. Precisely this kind of SM fields we shall consider
below. The exact expression for $f_1 (\varphi)$ is defined by the
form of the Lagrangian $L_m$. Obviously, if $\Omega_1$ is a
Liouville-type function, $f_1$ also belongs to this class of
functions. Corresponding Lagrangian density of interaction reads
\begin{eqnarray}\label{4-0}
\mathcal{L}_{int} &=& \eta \left. \frac{\delta\left(
\sqrt{h}F(\varphi ) L_m [h] \right)}{\delta
\varphi}\right|_{\varphi = \varphi_0} = \eta\left\{ \frac{d F}{d
\varphi} \sqrt{h} L_m [h] + F \frac{\partial h^{\mu \nu
}}{\partial \varphi} \frac{\delta \left( \sqrt{h} L_m [h]
\right)}{\delta h^{\mu \nu}} \right\}_{\varphi = \varphi_0}\nonumber \\
&=& \eta\left\{ \frac{d F}{d \varphi} \sqrt{h} L_m [h] - F
\frac{1}{\Omega_1}\frac{d \Omega_1}{d \varphi} \sqrt{h}
T^{\mu}_{\mu} [h] \right\}_{\varphi = \varphi_0}\, ,
\end{eqnarray}
where $T^{\mu}_{\mu} [h]$ is a trace of the energy-momentum tensor
for the matter Lagrangian $L_m [h]$ and $\Omega_1 (\varphi)$ is a
conformal factor connecting metrics in the string and Einstein
frames (see Eq. (\ref{2c})). It is worthy of note that Lagrangian
density $\sqrt{h} L_{em}[h]$ is invariant under conformal
transformation of metric for 4--D electromagnetic fields:
$\sqrt{\bar h} L_{em}[\bar h] = \sqrt{h} L_{em}[h]$ and $F
(\varphi) = f (\varphi)$. Thus, because a trace $T_{em} = 0$, the
Lagrangians of interactions (\ref{3b}) and (\ref{4-0})  formally
coincide with each other.


\vspace{0.5cm}

3. Minimal frame.

\vspace{0.5cm}

It can be easily seen, that there is additionally a specific frame
with a minimal coupling between dilaton and SM fields on the
brane. It corresponds to the conformal transformation
\begin{equation}\label{4a} \tilde g_{ab} = \Omega^2_2 (\varphi ) \bar
g_{ab}\, ,
\end{equation}
which yields
\begin{equation}\label{4b}
S_m = \int_{M_4} d^4 x \sqrt{|\tilde h|} L_m [\tilde h]  \, .
\end{equation}
To achieve it, the conformal factor $\Omega_2 (\varphi )$ should
satisfy the following condition
\begin{equation}\label{4c} \Omega^{-4}_2 (\varphi ) f (\varphi )
f_2 (\varphi ) = 1\, ,
\end{equation}
where, by the full analogy with function $f_1 (\varphi )$, the
exact expression for $f_2 (\varphi)$ is defined by the form of the
Lagrangian $L_m$. Obviously, for 4--D electromagnetic field $f_2 =
\Omega^4_2$ and equation (\ref{4c})  leads to $f (\varphi) \equiv
1$, i.e. dilaton should be minimally coupled with the brane
electromagnetic field from the very beginning. Thus, in the case
of non-minimal coupling, the transition to the minimal frame is
impossible for electromagnetic field.

In the minimal frame, the Lagrangian density of the interaction is
\begin{eqnarray}\label{4d}
\mathcal{L}_{int} &=& \eta \left. \frac{\delta\left( \sqrt{\tilde
h} L_m [\tilde h]\right)}{\delta \varphi}\right|_{\varphi =
\varphi_0} = \eta \left. \frac{\partial \tilde h^{\mu \nu
}}{\partial \varphi} \frac{\delta \left( \sqrt{\tilde h} L_m
[\tilde h] \right)}{\delta \tilde h^{\mu \nu}} \right|_{\varphi =
\varphi_0}\nonumber \\ &=& - \eta \left. \frac{1}{\Omega_2}\frac{d
\Omega_2}{d \varphi} \sqrt{\tilde h} T^{\mu}_{\mu} [\tilde h]
\right|_{\varphi = \varphi_0}\,
\end{eqnarray}
and has the form (\ref{2}) of interaction between radion and SM
fields. Therefore, in the minimal frame traceless fields do not
interact with dilaton at tree level. However, as we stressed
above, the minimal frame is absent for some of SM fields (if
original theory is non-minimal), e.g. for 4--D electromagnetic
field.

Now, we shall  concentrate on the experimental consequences of the
non-minimal coupling between dilaton and 4--D electromagnetic
field. In this case, transition to the minimal frame is absent and
the Lagrangian density of interaction has the same form in the
string as well as Einstein frames. In the following, to be more
concrete, we shall use the Einstein frame. In spite of the
traceless character of the electromagnetic field energy--momentum
tensor, Eqs. (\ref{3b}) and (\ref{4-0}) show that dilatonic fields
can interact on the brane with electromagnetic fields at the tree
level. It is the main difference with graviscalars considered in
Eq. (\ref{2}). Eq. (\ref{4-0}) can be rewritten as follows:
\begin{equation}\label{4} L_{int}
= \beta \frac{\psi}{M} L_m [h]= \left. \beta \frac{\psi}{M}
F_{\mu\nu}F^{\mu\nu}\right|_{h}\,  \end{equation}
with the coefficient $\beta := \left. d f/d \varphi
\right|_{\varphi_0}$. As we wrote above, $\varphi_0$ is the
present value of $\varphi$ and $\eta = M^{-1}\psi$ are small
fluctuations around it. Interaction (\ref{4}) is suppressed by the
electroweak mass\footnote{See also footnote 3 concerning 5--D ADD
models. For this approach $M \gtrsim 10^8$ GeV, which is much
bigger than 1TeV but is still much less than $M_{\mathrm{Pl}}$.}
$M=M_{\mathrm{EW}} \sim 1TeV$ in contrast to the interaction with
WIMP's (Weakly-Interacting Massive particles) which are suppressed
by 4-D Planck mass $M_{\mathrm{Pl}} \sim 10^{16}TeV$. Thus, the
interaction SM fields with dilatons in brane worlds can be much
more effective than with WIMP's in the standard Kaluza-Klein
approach.

Obviously, interactions between dilatons and massless SM
particles, e.g. photons,
are of great interest in high-energy colliding experiments. If the
dilaton field $\varphi$ is stabilized on the brane at $\varphi_0$
corresponding to a minimum of an effective potential and small
fluctuations near this position constitute quanta $\psi$ with a
mass $m$, then a decay rate (due to the interaction (\ref{4})) of
these quanta into 2 photons
are
\begin{equation}\label{5} \Gamma \sim \beta^2 \frac{m^3}{M^2}\,
\end{equation}
with a life-time $t \sim 1/\Gamma \sim
\beta^{-2}(M^2/m^3)(\hbar/c^2)$. Thus, the dilatons with masses
\begin{equation}\label{6} m \lesssim \beta^{-2/3}\left[
\frac{T_{\mathrm{Pl}}}{t_{\mathrm{univ}}}M^2M_{\mathrm{Pl}}\right]^{1/3}
\sim \beta^{-2/3}10^{-4}\mbox{eV} \end{equation}
have life-time $t \geq 10^{19}\mbox{sec} > t_{univ} \sim
10^{18}\mbox{sec} $ greater than the age of the Universe. They are
rather light particles. For heavier dilatons the decay plays
important role during the Universe evolution.

It is well known (see e.g. \cite{Carroll}) that interaction of the
form $f(\varphi ) F^{2}$ results in variation of the fine
structure constant $\alpha$ :
\begin{equation}\label{1d} \frac{\dot{\alpha}}{\alpha}\; =
\; \frac{\dot{f}}{f}\, ,
\end{equation}
where the dot denotes differentiation with respect to time. There
is an extensive list of papers devoted to the experimental bounds
for such variations (e.g. \cite{Hannestad,Webb,Melnikov} and
references therein). Different experiments give different bounds
on $|\dot{\alpha}/\alpha|$, from $\lesssim 10^{-12}\mbox{yr}^{-1}$
for cosmic microwave background \cite{Hannestad} to $\lesssim
10^{-17}\mbox{yr}^{-1}$ for the Oklo experiment \cite{DD}.
Primordial nucleosynthesis gives $|\Delta \alpha/\alpha| \lesssim
10^{-4}$ at a redshift on the orders $z = 10^9 - 10^{10}$
\cite{KPW}, i.e. $|\dot{\alpha}/\alpha| \lesssim
10^{-14}\mbox{yr}^{-1}$. In all these estimates $\dot{\alpha} =
\Delta \alpha /\Delta t$ is the average rate of change of $\alpha$
for the period $\Delta t$ (corresponding to a redshift $z$). For
our calculations we take some averaged estimate
$|\dot{\alpha}/\alpha| \lesssim 10^{-13}\mbox{yr}^{-1}$ which
corresponds to a Hubble time scale $\Delta t \sim H_0^{-1} \sim
10^{10}$ years. For this bound, from Eq. (\ref{1d}) we obtain:
\begin{equation}\label{2d}
\left| \frac{\dot{f}}{f}\right| = \left| \frac{\dot{\phi}}{M}
\frac{1}{f}\frac{d f}{d \varphi}\right| \lesssim
10^{-13}\mbox{yr}^{-1} \, ,
\end{equation}
This estimate leads to the following restriction on the parameter
$\beta$ (cf. \cite{Carroll}):
\begin{equation}\label{7} |\beta|\approx \Delta t \left|
\frac{\dot{\alpha}}{\alpha}\right| \frac{M}{\Delta \phi} \;
\Rightarrow \; |\beta| \lesssim 10^{-3}\, ,
\end{equation}
where we suppose $\Delta \phi \sim M$ and that the present value
of $f \approx 1$ (that usualy is equivalent to the assumption for
the dilaton field at the present time: $\phi_0 \ll M \Rightarrow
\varphi_0 \ll 1$).


As we wrote above, most of the dilatonic models are motivated by
string theories
which, at a low-energy limit, usually have the Liouville-type
potentials (see e.g. Eq. (\ref{3})): $V(\varphi )= V \exp
(a\varphi )$ and $f(\varphi ) = \exp (b\varphi )$ with $a \sim b
\sim \mathcal{O}(1)$ \cite{Lidsey,OS,GNS,BDB}. Substitution of the
Liouville coupling potential $f(\varphi ) = \exp (b\varphi )$ into
estimate (\ref{2d}) leads to the limits on the parameter $b$:
\begin{equation}\label{8} \left| b \frac{1}{M} \frac{\Delta \phi}
{\Delta t}\right| \lesssim 10^{-13}\mbox{yr}^{-1} \; \Rightarrow
\; |b| \lesssim 10^{-3}\,
\end{equation}
for $\Delta \phi \sim M$ and $\Delta t \sim 10^{10}$ years.
Estimates (\ref{7}) and (\ref{8}) coincide with each other because
for $\varphi_0 \ll 1: \beta = \left. d f/d \varphi
\right|_{\varphi_0} = b \exp (b \varphi_0) \sim b$. It is worth
pointing out that natural assumption $\Delta \phi \sim M$ results
in independence of estimates (\ref{7}) and (\ref{8}) upon the
concrete value of $M$.

It is hardly possible that the Liouville-type potentials for such
considered model provide the stabilization of $\varphi$ on the
brane.
Thus, the dilatonic models with
non-minimal coupling to the SM fields on the brane are ruled out
by estimate (\ref{8}) for theories with $a \sim b \sim
\mathcal{O}(1)$ (e.g. model (\ref{2b}), (\ref{3})).

Another restrictions on parameter of models can be obtained from
experiments on variation of the 4--D gravitational constant. To
show it, we represent the brane part of an action (in the Einstein
frame) as follows:
\begin{equation}\label{a.0} S_b = \int_{M_4}
d^4 x \sqrt{|h|} \{ - T(\varphi ) + F( \varphi)L_m [h] \} \, ,
\end{equation}
where $T(\varphi ) \equiv \Omega_1^{-4} (\varphi ) f( \varphi )
T$. For 4--D electromagnetic field $F(\varphi ) = f(\varphi )$ and
we shall use this equality below.

In 4--D projected Einstein equations (see e.g. \cite{MW,MB}),
linear terms (with respect to the brane matter energy-momentum
tensor\footnote{An effective energy-momentum conservation equation
for the matter on the brane has the form \cite{MB} $( f(\varphi)
T^{\nu}_{\mu}[h])_{; \nu} = \varphi_{,\mu } (d f/d \varphi) L_m[h]
\, ,$ which shows that the matter is conserved on the brane if the
dilaton field is either minimally coupled to the SM ($f \equiv$
const) or stabilized on the brane ($\left. \varphi\right|_{brane}
\rightarrow $ const).} $T_{\mu \nu}[h] = -2 \delta L_m[h] / \delta
h^{\mu \nu} + h_{\mu \nu} L_m[h]$) are responsible for the
conventional cosmology on the brane. For the brane-world models
with $S_b$ of the form (\ref{a.0}), these linear terms have the
form $(\kappa^4_5 /6) T(\varphi) f(\varphi) T_{\mu \nu}[h]$. Thus,
the quantity
\begin{equation}\label{a.5}  8 \pi G_N \equiv \frac{\kappa^4_5}{6}
T(\varphi ) f(\varphi ) \equiv \frac{\kappa^4_5}{6} \tilde
f(\varphi )\end{equation}
plays the role of a 4-D gravitational constant on the brane.
This equation shows that effective 4--D gravitational constant
$G_N$ for models (\ref{a.0}) defines by the function $\tilde
f(\varphi)$. Thus, variation of $\tilde f(\varphi)$ leads to a
variation of $G_N$. If functions $\Omega_1(\varphi)$ and
$f(\varphi)$ are of the Liouville-type then $\tilde f (\varphi)$
also belongs to this class of functions and can be written in the
form $\tilde f (\varphi) = T \exp (c\, \varphi )$. Obviously, for
models with $a \sim b \sim \mathcal{O}(1)$ parameter $c \sim
\mathcal{O}(1)$. For example, for model (\ref{2b}), (\ref{3}) we
have: $T(\varphi) = T \exp\left( (2/3)\varphi\right)\, , \;
f(\varphi) = \exp (-2\varphi)$ and $\tilde f(\varphi) = T
\exp\left( (-4/3)\varphi\right)$.

There are a number of observable data for an estimate of a
possible time variation of the gravitational constant
\cite{Melnikov,DFRRW,Damour,Will}. They imply $|\dot{G}_N/G_N|
\lesssim 10^{-11}\mbox{yr}^{-1}$. Thus, we can obtain a limitation
of the variation of $\tilde f(\varphi)$:
\begin{equation}\label{9} \left| \frac{\dot{\tilde f}}{\tilde{f}} \right| \lesssim
10^{-11}\mbox{yr}^{-1}\, ,\end{equation}
which for the Liouville potential $\tilde f = T\exp (c\, \varphi)$
puts on the parameter $c$ the following restrictions:
\begin{equation}\label{10} |c|\lesssim 10^{-1}\, , \end{equation}
where we suppose $\Delta \phi/\Delta t \sim M/(10^{10}\mbox{yr})$.
This estimate is much less severe than (\ref{8}) and, strictly
speaking, not rules out theories with $c \sim \mathcal{O}(1)$. For
example, it is expected \cite{Damour,Will}, that on the Hubble
time scale $|\dot{G}_N/G_N| \lesssim H_0 \sim
10^{-10}\mbox{yr}^{-1}$. Then, inequality (\ref{10}) is reduced to
the following estimate: $|c| \lesssim 1$.

In order to avoid the problem of the fundamental constant
variation in the non-minimal dilatonic brane-world models, it is
natural to suppose that the dilaton is stabilized on the brane
(before primordial nucleosynthesis), i.e. $\varphi \rightarrow
\varphi_0 \equiv$ const where $\varphi_0$ corresponds to a stable
solution of the equation of motion on the brane. It would be very
interesting to find explicit models leading to such stabilization.
If, in general, such stabilization is impossible, then, variations
of $\varphi $ with time should be in accordance with experimental
bounds on variations of the fundamental constants (see e.g. Eqs.
(\ref{2d}) and (\ref{9})).

\vspace{1cm}


{\bf Acknowledgements}

I would like to thank Valery Rubakov for interesting discussions,
Paulo Moniz for useful comments and Uwe G\" unther for a kind
assistance with the literature. I also acknowledge support by the
programme SCOPES (Scientific co-operation between Eastern Europe
and Switzerland) of the Swiss National Science Foundation, project
No. 7SUPJ062239.

%
%

%
%
\end{document}